# Information and Errors in Quantum Teleportation

Balaji Nedurumalli

## Introduction

Quantum teleportation [1-3] is a striking illustration of the difference between classical and quantum communication processes [4]. In teleportation, communication is facilitated by entanglement. But state transmission is predicated on belief that the entanglement between the resource particles is perfect and the protocol steps are error-free, both of which it is impossible to establish unconditionally.

The dictionary definition of teleportation is generally "transportation of matter through space by converting it into energy and then reconverting it at the terminal point." Thus, according to the Free Online Dictionary, the meaning is "transport by dematerializing at one point and assembling at another." Quantum teleportation does not live up to this meaning of the term. The teleportation process only transfers the "information" related to the "unknown" observable to a remote location using the resource of two entangled particles. In such a conception, the unknown information could be teleported to a different kind of particle, if the entangled particle resource is different from that of the particle whose state is sought to be transported. This is what happens in the experiments where the state of an atom is teleported to a remote location [5].

In this article, we consider the question of the teleportation protocol from an engineering perspective. Suppose the experimental arrangement to perform teleportation exists, what can one say regarding the fidelity of the process? Also, how do we view the information process?

## The teleportation protocol

For background, we examine the teleportation protocol (e.g. [6]) to investigate the assumptions that must be fulfilled for successful teleportation to take place. Let the orthogonal Bell state $|B_1>$ be represented as:

$$|B_1> = |00> + |11>$$

in which we leave out the normalizing factor for the probability amplitudes for convenience.

The teleportation process is pictured in Figure 1, and it requires the resource of the two entangled particles $|B_1>$, one of which is with Alice and the other is with Bob. But how do we know that the two particles are perfectly entangled to form the Bell state $|B_1>$? These particles cannot be checked to see if they satisfy the entanglement condition, because such a checking would make their states collapse. The teleportation protocol uses elementary operations on the qubits.

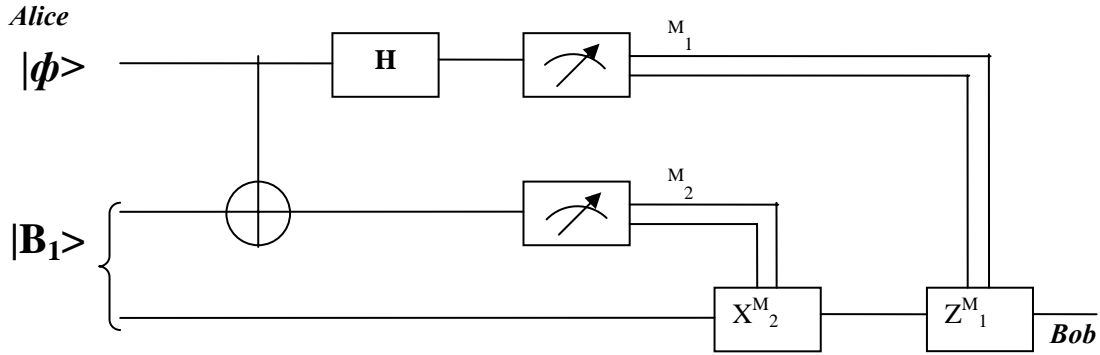

**Figure 1:** The standard teleportation protocol

We must, therefore, assume that Alice and Bob take it on faith that the entangled pair $|B_1\rangle$ is error-free. This is a major assumption with implications for the information process.

The steps of the teleportation protocol are straightforward:

1. Alice sends an unknown qubit $|\phi\rangle = a\,|0\rangle + b\,|1\rangle$ with unknown $a, b$. The initial state of the three qubits would be

$$a\,|000\rangle + b\,|100\rangle + a\,|011\rangle + b\,|111\rangle$$

2. An XOR gate is applied between the first two qubits at the transmitter (Alice) which yields the state:

$$a\,|000\rangle + b\,|110\rangle + a\,|011\rangle + b\,|101\rangle$$

3. Alice applies a *H* operator to the first qubit to yield the state

$$a\,(|000\rangle + |100\rangle) + b\,(|010\rangle - |110\rangle) + a\,(|011\rangle + |111\rangle) + b\,(|001\rangle - |101\rangle)$$

The above state may be rewritten as:

$$|00\rangle(a|0\rangle + b|1\rangle) + |01\rangle(a|1\rangle + b|0\rangle) + |10\rangle(a|0\rangle - b|1\rangle) + |11\rangle(a|1\rangle - b|0\rangle)$$

4. Alice measures the first two qubits and the remaining qubits collapse to one of the four states below

$$a|0\rangle \pm b|1\rangle \;,\; a|1\rangle \pm b|0\rangle$$

5. The information on the first two qubits measured by Alice is now sent to Bob through a classical channel such that Bob can decode the remaining qubits using the appropriate operators given below and place them in the state $|\phi\rangle$. The operators used by Bob for the corresponding measurements made by Alice are given below:

$$\begin{bmatrix} 1 & \\ & 1 \end{bmatrix} \quad \text{for} \quad |00\rangle$$

$$\begin{bmatrix} & 1 \\ 1 & \end{bmatrix} \quad \text{for} \quad |10\rangle$$

$$\begin{bmatrix} 1 & \\ & -1 \end{bmatrix} \quad \text{for} \quad |01\rangle$$

$$\begin{bmatrix} & -1 \\ 1 & \end{bmatrix} \quad \text{for} \quad |11\rangle$$

The entangled qubit with Bob comes to have the original state of the unknown qubit after it has been operated upon as shown above.

## Imprecision and error

Let's examine the engineering requirements for the various steps of the protocol. For the unknown quantum state to reappear with complete fidelity at the remote location, the following resources are essential:

1. The entangled pair $|B_1\rangle$ should be completely isolated from other particles. Furthermore, the entanglement should be error-free.

2. There should be no errors in the performance of the XOR operation by Alice.

3. The Hadamard operator $H$ should be applied with complete precision.

4. The operators applied by Bob upon his entangled particle, after he has received classical information from Alice, should be error-free.

None of these requirements can be met absolutely. The operators are controlled by classical controls, and they would suffer from errors [7-8]. Also, the very control of an XOR gate is not guaranteed to be error-free [9].

The random errors in the various steps of the process cannot be taken to be linear, although the gate operations in themselves are linear, because of the nonlinear nature of the measurement process. This means that the error in the teleported state with Bob can be much greater than the errors in the components.

The analysis of the teleportation protocol shows that it should not be seen as a consequence of nonlocality [10-12]. The use of the mental picture related to the consideration of an isolated system whose informational properties are being sought is inadequate [13-16], which explains the difficulties that have been encountered in the implementation of practical quantum computing systems [17-19].

## Certification of the Entangled Pair

The teleportation protocol would ideally require an authority that ensures that the particle pair that is furnished to Alice and Bob is indeed entangled. But there is no way this can be established to the satisfaction of Alice and Bob due to the fact that an unknown quantum state cannot be copied. But this sets up the requirement of a chain of authorities, one sitting on top of another.

This is further indication that the foundations of quantum information cannot rest on the same basis as that of classical information [14]. In classical communication, there is a certain expectation of what the state to be received is going to be, and once it is received there can be a protocol set up to determine if the expectation turned out to be correct. This is possible because classical states can be copied.

If there is no proof to be furnished that the state received by Bob is the one that was originally with Alice, then what do we mean by the assertion that the unknown state has been teleported? This means that as argued by Kak [4], quantum information cannot be defined absolutely, and that it should be represented in terms of the observer.

## Conclusions

The notion that information is being "teleported" is perhaps not quite adequate to communicate what is happening. The state, together with the entangled pair of particles, already forms one large system. Given that, all that is happening is the transfer of the information from one part to another, within the system.

Though quantum physics shows the theoretical possibility for teleportation, its practical implementation in a manner that has absolutely no error is not feasible. There would be inevitable uncertainty in the working of the components of the protocol, which would get reflected in uncertainty related to the state that is transferred to Bob. In this practical sense, it is not possible to teleport an unknown state.

# References

1. C. H. Bennett, G. Brassard, C. Crepeau, R. Jozsa, A. Peres, and W. Wootters, 1993. Teleporting an unknown state via dual classical and EPR channels. Phys. Rev. Lett., 70, pp. 1895–99.
2. N.D. Mermin, 2001. From Classical State-Swapping to Teleportation. Phys. Rev. A, 65, p. 012 320; arXiv: quant-ph/0105117.
3. S. Kak, 2003. Teleportation protocols requiring only one classical bit. arXiv: quant-ph/0305085.
4. S. Kak, 1998. Quantum information in a distributed apparatus. Found. Phys. 28: 1005-1012; arXiv: quant-ph/9804047.
5. K. Hammerer, E.S. Polzik, J.I. Cirac, 2005. Teleportation and spin squeezing utilizing multimode entanglement of light with atoms. Phys. Rev. A 72, 052313.
6. S. Kak, 2003. Bell states, dense coding, and teleportation. arXiv:quant-ph/0304184
7. S. Kak, 2000. Rotating a qubit. Inform. Sc., 128, pp. 149-154. arXiv: quant-ph/9910107.
8. S. Kak, 2002. Uncertainty in quantum computation. arXiv: quant-ph/0206006.
9. S. Kak, 2006. Information complexity of quantum gates. Int. J. Theo. Phys. 45; arXiv: quant-ph/0506013.
10. L. Hardy, 1999. Disentangling nonlocality and teleportation. arXiv:quant-ph/9906123.
11. J.S. Barrett, 2001. Implications of teleportation for nonlocality. Physical Review A, 64, p. 042 305; arXiv:quant-ph/0103105.
12. R. Penrose, 1998. Quantum computation, entanglement and state reduction. Philosophical Transactions of the Royal Society of London A, 356, pp. 1927–39.
13. S. Kak, 2001. Statistical constraints on state preparation for a quantum computer. Pramana 57, pp. 683-688; arXiv: quant-ph/0010109.
14. C. Brukner and A. Zeilinger, 2001. Conceptual inadequacy of the Shannon information in quantum measurements. Physical Review A, 63, p. 022 113.
15. S. Kak, 2003. General qubit errors cannot be corrected. Inform. Sc. 152, 195-202; arXiv: quant-ph/0206144.
16. S. Kak, 1999. The initialization problem in quantum computing. Found. Phys., 29: 267-279; arXiv: quant-ph/9805002.
17. S. Kak, 2001. Are quantum computing models realistic? arXiv: quant-ph/0110040.
18. S. Kak, 2003. Representation of entangled states. arXiv: quant-ph/0302118
19. A. Ponnath, 2006. Difficulties in the implementation of quantum computers. arXiv: cs.AR/0602096.
20. M.A. Nielsen and I.L. Chuang, 2000. Quantum Computation and Quantum Information. Cambridge University Press.